\documentclass[12pt,preprint]{aastex}
\usepackage{apjfonts}
\def\beq{\begin{equation}}
\def\enq{\end{equation}}
\def\bea{\begin{array}}
\def\ena{\end{array}}
\def\msun{M_\odot}
\def\fdg{\hbox{$.\!\!^\circ$}}

\newcommand{\siml}{\lower4pt \hbox{$\buildrel < \over \sim$}}
\newcommand{\simg}{\lower4pt \hbox{$\buildrel > \over \sim$}}
\bibpunct[,]{(}{)}{;}{a}{}{,}

\begin{document} 
\title{\sc The Spin of the Black Hole in the Soft X-ray Transient
A0620--00}

\shorttitle{The spin determination for A0620--00}
\shortauthors{Gou et al.}


\author{Lijun Gou$^{1}$, Jeffrey E.\ McClintock$^{1}$,
James F.\ Steiner$^{1}$, Ramesh Narayan$^{1}$, Andrew G.\
Cantrell$^{2}$, Charles D.\ Bailyn$^{2}$, Jerome A.\ Orosz$^{3}$}
\altaffiltext{1}{Harvard-Smithsonian Center for Astrophysics,
  Cambridge, MA, 02138}
\altaffiltext{2}{Department of Astronomy, Yale University, PO Box
  208101, New Haven, CT 06520}
\altaffiltext{3}{Department of Astronomy, San Diego State University,
  5500 Campanile Drive, San Diego, CA 92182}

\slugcomment{ApJL}
\begin{abstract}
During its yearlong outburst in 1975--76, the transient source
A0620--00 reached an intensity of 50 Crab, an all-time record for any
X-ray binary.  The source has been quiescent since.  We recently
determined accurate values for the black hole mass, orbital
inclination angle and distance. Building on these results, we have
measured the radius of the inner edge of the accretion disk around the
black hole primary by fitting its thermal continuum spectrum to our
version of the relativistic Novikov-Thorne thin-disk model.  We have
thereby estimated the spin of the black hole.  Although our spin
estimate depends on a single high quality spectrum, which was obtained
in 1975 by {\it OSO-8}, we are confident of our result because of the
consistent values of the inner-disk radius that we have obtained for
hundreds of observations of other sources: H1743-322, XTE J1550-564,
and notably LMC X-3.  We have determined the dimensionless spin
parameter of the black hole to be $a_*=0.12 \pm 0.19$, with
$a_*<0.49$ and $a_*>-0.59$ at the $3\sigma$ level of confidence.  This
result takes into account all sources of observational and
model-parameter uncertainties.  Despite the low spin, the intensity
and properties of the radio counterpart, both in outburst and
quiescence, attest to the presence of a strong jet.  If jets are
driven by black hole spin, then current models indicate that jet power
should be a steeply increasing function of $a_*$.  Consequently, the
low spin of A0620--00 suggests that its jet may be disk-driven.

\end{abstract}


\section{\sc Introduction}

A0620--00 is the prototype soft X-ray transient (SXT), which is an
eruptive type of X-ray binary.  For several days in 1975 the flux at
Earth from this source was greater than the combined total flux of all
the other Galactic X-ray binaries, including Sco X-1.  A decade later,
a dynamical study of its quiescent optical counterpart (V616 Mon) led
to the discovery of the first black hole (BH) primary in an SXT
\citep{mr+1986}.  A0620--00 is one of eight, similar short-period BH
SXTs ($P_{\rm orb}<12$~hr; \citeauthor{rm+06}~\citeyear{rm+06}).  In
the optical band, it is the best-studied of these systems because its
counterpart is bright ($V_{\rm quiescent} = 18.3$) and close: $D =
1.06\pm0.12$ kpc (\citeauthor{cbom+09}~\citeyear{cbom+09}).

The radial velocity amplitude of the secondary star is firmly
established and has now been determined to the remarkable precision of
0.1\% \citep{nsv+08}.  However, a reliable estimate of the BH mass
$M$, which depends on a robust determination of the orbital
inclination angle $i$, has remained elusive.  Several measurements of
$i$ have been made by modeling the ellipsoidal variability of the
secondary, but they have been inconsistent \citep{nsv+08}.  The
determination of $i$ is complicated by a variable and phase-dependent
component of light from the accretion disk.  Recently, however, a
comprehensive analysis of all of the available light curve data (32
data sets spanning 30 years) points to consistent values of
inclination and BH mass: $i=51\fdg0\pm0\fdg9$~and
$M=6.61\pm0.25~M_{\odot}$ \citep{cbom+09}.

Our group has published spin estimates for five stellar-mass BHs
\citep{smnd+06,msnr+06,lmnd+08,gmln+09}.  Meanwhile, the spins of
several stellar-mass BHs have also been obtained by modeling the
profile of the Fe K line (see \citeauthor{mrfm+09}~\citeyear{mrfm+09},
and references therein).  The spins we find are all quite high, with
values of the spin parameter $a_*$ in the range 0.7 to $>0.98$.  The
dimensionless quantity $a_* \equiv cJ/GM^2$ with $|a_*| \le 1$, where
$M$ and $J$ are respectively the BH mass and angular momentum
\citep{st+1986}.  We use the X-ray continuum-fitting method,
which was pioneered by \citet{zcc+97}.  Our spin estimates are based
on our version of the Novikov-Thorne thin accretion disk model
\citep{lznm+05} and an advanced treatment of spectral hardening
\citep{dbht+05,dh+06}.  We only consider spectra that contain a
dominant thermal component \citep{smrn+09} and for which the
Eddington-scaled bolometric disk luminosity is moderate, $l\equiv
L_{\rm bol}(a_*, \dot{M})/L_{\rm Edd}<0.3$ \citep{msnr+06}.

For the continuum-fitting method to succeed it is essential to have
accurate values of $M$, $i$ and $D$ \citep[e.g.,][]{gmln+09}, such as
those reported above for A0620--00.  Herein, we use these input data
and the only suitable, extant X-ray spectrum of the source in order to
estimate the spin of the BH primary.  Although our spin measurement is
based on a single spectrum, we have considerable confidence in our
result for the following two reasons: (1) In our earlier work, we have
found that our measurements of spin, or equivalently the dimensionless
radius of the innermost stable circular orbit $r_{\rm isco}\equiv
R_{\rm ISCO}/(GM/c^2)$), for a particular source are remarkably
consistent over years or decades
\citep[e.g.,][]{gmln+09,smrn+09,smrn+092}.  Most compelling is our
recent study of the persistent source LMC X-3: Hundreds of
observations obtained over a span of 26 years by eight different
missions give values of $r_{\rm isco}$, and hence $a_*$, that are
consistent within a few percent (Steiner et al.\ 2010).  One important
conclusion from this study is that a single high quality spectrum is a
good proxy for a large collection of spectra.  (2) The {\it OSO-8}
spectrum in question is a very high quality spectrum for the
determination of spin via the continuum-fitting method: It is a
remarkably pure thermal spectrum that is almost completely free of the
effects of Comptonization (Section 3), and it was obtained using a
stable, advanced proportional-counter detector that was the forerunner
of the {\it RXTE} PCA detectors.

\section{\sc Data Selection and Reduction}

A0620--00 was observed using the Goddard Cosmic X-ray Spectroscopy
Experiment (GCXSE) aboard {\it OSO-8} \citep{sbbh+76,ssbh+77}.  The
detector employed, namely the C Detector, is a sealed multiwire
xenon-methane proportional counter with a net effective area of
237~cm$^2$, which is fitted with a 5\fdg1 (FWHM) circular collimator.
The observation commenced on 1975 September 29 at 11:58~UT and
continued for 3.0 days during which the intensity of the source was
$\approx11$ Crab \citep{mbbc+76}.  The average collimator-corrected
raw count rate was $\approx1650$ counts~s$^{-1}$ and varied by
$\approx20$\% over the course of the observations.  The gain and
detector resolution (18\% at 6 keV) were determined using an on-board
$\rm ^{241}Am$ source, and the energy calibration is good to a
precision of $\sim1\%$ \citep{msbh+78}.

We first ran the FTOOL {\it osofindfast} to determine the ``good
observing days'' and then downloaded the appropriate raw PHA data.
These data had already been filtered to eliminate times of high
background during passage through the South Atlantic Anomaly and times
of Earth occultation.  Attending only to the data from Detector C, we
ran the tool {\it osopha} to extract the 63-channel spectrum (spanning
2--60 keV), and we added a 2\% systematic error to all the channels;
{\it osopha} automatically corrected the counting rates for dead time
by the factor 0.704.  The effective source exposure time for the
observation is 21.3 ks.  We then used the tools {\it osofindfast} and
{\it osopha} to extract and examine background spectra from several
locations. Our results are insensitive to the choice of the background
region, and we finally settled on a region centered at $l=201\fdg9$
and $b=-28\fdg2$ and an effective background observation time of 3.35
ks.  We computed a correction to the response of the detector by
comparing the power-law spectrum of \citet{ts+1974}, our standard
reference spectrum \citep{msnr+06}, to the parameters derived by
analyzing a spectrum of the Crab that was obtained just 12 days prior
to the observation of A0620--00.  Following a new correction procedure
\citep{smrn+092}, we computed a \citeauthor{ts+1974} normalization
coefficient $f_{\rm TS}=0.792$ and a slope difference
$\Delta{\Gamma_{\rm TS}}=-0.148$, where $f_{\rm TS}$ is the ratio of
the observed normalization to that of \citeauthor{ts+1974}, and
$\Delta{\Gamma_{\rm TS}}$ is the difference between the observed value
of the photon index and that of the reference spectrum.  This
correction is applied in all the analysis work below via a custom
XSPEC multiplicative model.

\section{\sc Analysis and Results}

{\it Preliminary, nonrelativistic analysis:} All of the data analysis
and model fitting throughout this paper were performed using XSPEC
version 12.6 \citep{arna+96}.  As in our earlier work
\citep[e.g.,][]{gmln+09}, we first make an assessment of the data by
analyzing it with the nonrelativistic disk model {\sc diskbb}.  For
modeling the weak Compton component of emission we use our convolution
model {\sc simpl} \citep{snme+09}, which far outperforms the standard
power-law model ({\sc powerlaw}) with its troublesome divergence at
low energies \citep{twz+05,smrn+09}.  The parameters of {\sc simpl}
and {\sc powerlaw} are similar; their principal parameter, the photon
index $\Gamma$ is identical.  However, the normalization parameter for
{\sc simpl} is the scattering fraction -- the fraction $f_{\rm SC}$ of
the seed photons that are scattered into the power-law tail -- rather
than the photon flux.

Thus, the model we first employ is {\sc tbabs(simpl$\otimes$diskbb)},
where {\sc tbabs} is a widely-used model of low-energy absorption
\citep{wam+00}. We fitted the data over the energy range 2.2--17 keV
\citep{msbh+78}.  Because of the detector's limited low-energy
response, we are unable to fit for the hydrogen column density $N_{\rm
H}$, which we estimate from published values of the reddening.  Based
on the five refereed papers known to us, we find $0.25<E(B-V)<0.45$
with a most frequently-cited value of 0.35 \citep{wphs+83}.  We adopt
the value $E(B-V)=0.35\pm0.05$.  Assuming $A_{\rm V}/E(B-V)=3.1$ and
$N_{\rm H}/A_{\rm V}=2.0\times10^{21}$~$\rm mag^{-1}$~cm$^{-2}$
\citep{ps+1995, go+0903}, we obtain our estimate of the column
density: $N_{\rm H}=2.2\times 10^{21}~{\rm cm^{-2}}$.

The fit to the PHA spectrum is good ($\chi_{\rm \nu}^{2}=1.03$), and
the temperature is precisely determined: $kT=0.700\pm0.004$~keV.
Moreover, the parameters of the Comptonized emission are well
determined: $\Gamma=3.81\pm0.37$ and $f_{\rm SC}=0.008\pm0.003$.  The
best-fit value of the scattering fraction is low, 0.8\% (compare
\citeauthor{gmln+09}~\citeyear{gmln+09} and
\citeauthor{smrn+09}~\citeyear{smrn+09}), and it is even lower for the
relativistic model (0.6\%; see below).  Thus, this spectrum is thermal
dominant in the extreme \citep{rm+06}.  Because the Compton component
is faint the fitted values of both $kT$ and $a_*$ depend very weakly
on how one models this nonthermal emission.

We are fortunate that in porting the {\it OSO-8} GCXSE software/data
to the HEASARC that precisely the same spectrum of A0620--00 we
consider was used to illustrate how one reduces and analyzes {\it
OSO-8}
data\footnote{http://heasarc.gsfc.nasa.gov/docs/oso8/software/oso8\_example3.html}.
The model employed in this example is {\sc phabs(diskbb+powerlaw)},
and the fixed column density is $\approx35$\% higher, $N_{\rm
H}=3.0\times 10^{21}~{\rm cm^{-2}}$.  Using this model, we obtain
precisely the same temperature reported in the example, namely
$kT=0.70$~keV, which confirms the correctness of our
reduction/analysis procedures.


{\it Relativistic analysis:} We turn now to the analysis of the data
using our fully relativistic accretion disk model {\sc kerrbb{\small
2}}, which includes self-irradiation of the disk (``returning
radiation'') and limb darkening \citep{lznm+05}.  The effects of
spectral hardening are incorporated into the basic model {\sc kerrbb}
via a pair of look-up tables for the hardening factor $f$
corresponding to two representative values of the viscosity parameter:
$\alpha=0.01$ and 0.1 \citep[see][]{gmln+09}.  The entries in this
table were computed using a second relativistic disk model {\sc
bhspec} \citep{dbht+05}.  We refer to the model {\sc kerrbb} plus this
table/subroutine as {\sc kerrbb{\small 2}}.  The model {\sc
kerrbb{\small 2}} has just two fit parameters, namely the BH spin
parameter $a_*$ and the mass accretion rate $\dot M$.  For further
details see \citet{msnr+06}.

We now analyze the data in exactly the same manner as before, except
that we replace {\sc diskbb} by {\sc kerrbb{\small 2}}: {\sc
tbabs(simpl$\otimes$kerrbb{\small 2}}).  We fix the column density at
$N_{\rm H}=2.2\times 10^{21}~$cm$^{-2}$, and we have four free
parameters: the dimensionless spin parameter $a_*$, the mass accretion
rate $\dot{M}$, the photon index of the high energy component
$\Gamma$, and the scattering fraction $f_{\rm SC}$.  We fitted the
spectrum with the input parameters fixed at their baseline values (see
footnote to Table~\ref{tablefirst}).  The normalization was fixed at
unity (as appropriate when $M$, $i$ and $D$ are held fixed).  We
included the effects of limb darkening (lflag = 1) and returning
radiation (rflag = 1), and we set the torque at the inner boundary of
the disk to zero ($\eta=0$).  The fitted results are presented in the
first row of Table~\ref{tablefirst}.  Compared to the nonrelativistic
fit, the fit here is even better ($\chi_{\rm \nu}^{2}=0.74$).  The
luminosity is moderate ($l\approx0.1$) and easily meets our selection
criterion $l <0.3$ (Section\ 1).  Of prime interest, we find a low and
precise value of the spin parameter, $a_*=0.135\pm0.029$.

\begin{figure}
\plotone{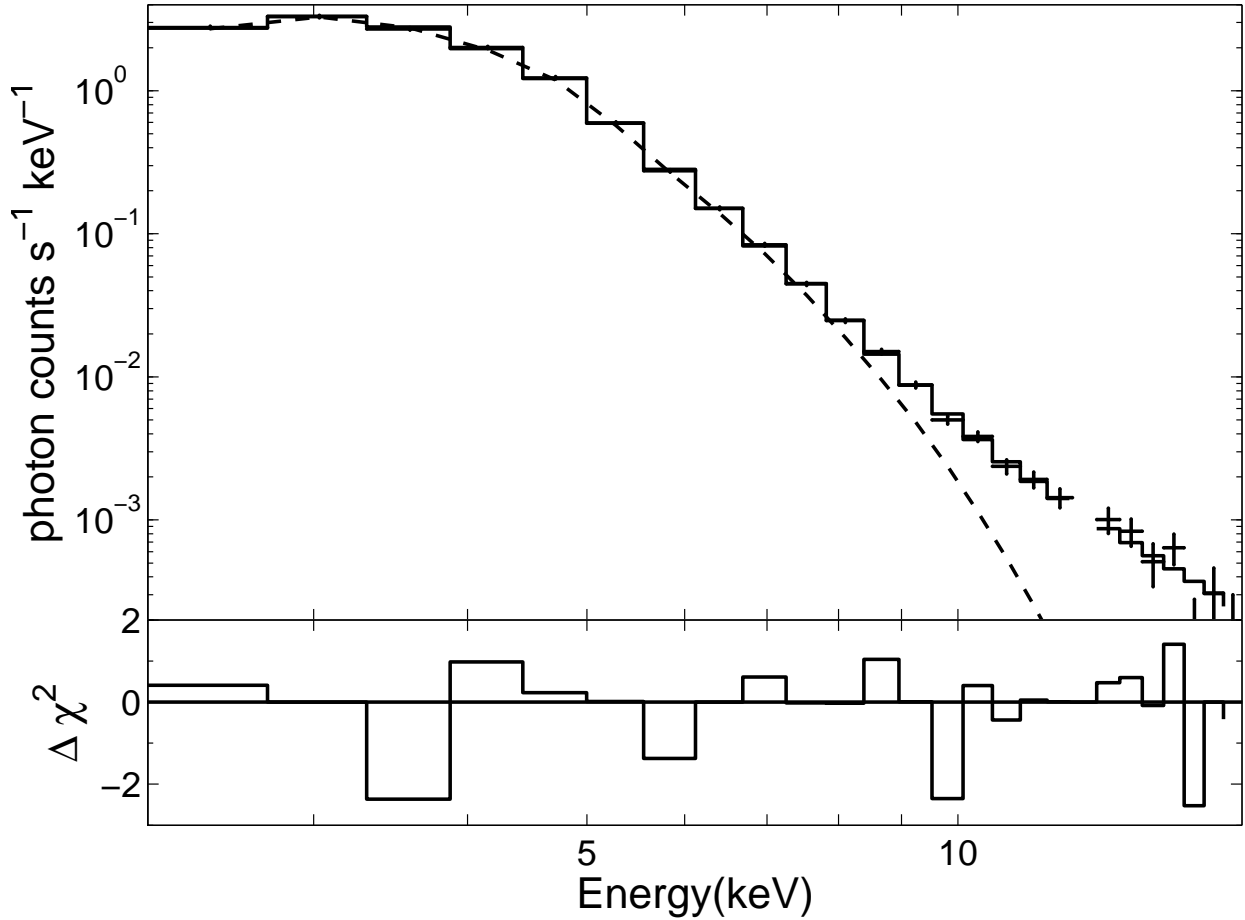}
\caption{Unfolded {\it OSO-8} X-ray spectrum of A0620--00.  {\it Top:}
  The histogram shows the model fitted to the data with the dashed
  line representing the thermal component. {\it Bottom:} The residuals
(data minus model in units of $\chi_{\rm \nu}^{2}$).  The data for the bin
  centered at $E\approx12.7$~keV were corrupted and are ignored (see
  footnote 1 in the text).}
\label{figurefirst}
  \end{figure}

The unfolded photon spectrum and fit residuals are shown in
Figure~\ref{figurefirst}.  The best-fit value of the scattered
fraction is very low, only 0.6\% (Table~\ref{tablefirst}).  The second
and third rows of Table~\ref{tablefirst} show that the uncertainty in
$N_{\rm H}$ has a quite modest effect on the value of the spin
parameter, shifting the best-fit value of $a_*$ by $<0.22\sigma$.  The
last two rows of Table~\ref{tablefirst} further show that fixing the
power-law index $\Gamma$ at $\pm1{\sigma}$ from its best-fit value of
3.53 also has a modest effect on $a_*$, shifting its best-fit value by
at most $\approx0.55\sigma$.  In fact, relative to the {\it total}
uncertainty in $a_*$ these small shifts are over ten times smaller
still, i.e., $\siml 0.1\sigma$, as we now show.


{\it Comprehensive error analysis:} The statistical uncertainty in
$a_*$ is small, and other sources of error dominate.  Ignoring for now
uncertainties in the theoretical model (see Section 4), we consider
the effects of uncertainties in (1) the input parameters $M$, $i$ and
$D$, (2) the column density $N_{\rm H}$, (3) the viscosity parameter
$\alpha$ \citep{dbht+05}, and (4) the metallicity $Z$.  We first note
that, as in the case of LMC X-1 \citep{gmln+09}, the uncertainty in
the metallicity of the disk gas is negligible compared to the
statistical error : We find that $a_*$ changes by only $0.34\sigma$
when we vary the metallicity from our default value of $Z=1$ (solar)
to $Z=0.1$.

In order to determine the error in $a_*$, we performed an analysis
that considers at once the above items (1) and (2), i.e., the combined
uncertainties in $M$, $i$, $D$ and $N_{\rm H}$.  We fix the viscosity
parameter to its baseline value, $\alpha=0.01$.  In order to determine
the error in $a_*$, we performed Monte Carlo simulations assuming that
the uncertainties in the four parameters are normally and
independently distributed.  Specifically, we (1) generated 3000
parameter sets for $M$, $i$, $D$ and $N_{\rm H}$; (2) computed for
each set the look-up table for the spectral hardening factor $f$ using
the model {\sc bhspec}; and (3) using these $f$-tables, obtained $a_*$
by fitting our model to the spectrum.  The results for the 3000
simulation runs are shown in Figure~\ref{figuresecond}.  The
uncertainty in $D$ (panel {\it c}) dominates the error in $a_*$, with
the uncertainty in $M$ having less than half as much effect, while the
uncertainties in $i$ and $N_{\rm H}$ are smaller still.  The
corresponding histogram for the spin displacements in $a_*$ about the
most probable value is shown in Figure~\ref{figurethird} by the light
solid line.

The analysis above is for our baseline value of $\alpha=0.01$.
Following precisely the same procedures, we also performed the Monte
Carlo analysis for $\alpha=0.1$, and the result is the dashed-line histogram
shown in Figure~\ref{figurethird}.  The summation of these two
histograms results in the large histogram (heavy solid line).  The
combined distribution, which corresponds to a total of 6000
simulations, has a median spin value of 0.12 and implies a $1\sigma$
error of (-0.19, +0.19).  {\it Thus, considering all significant
observational and model-parameter uncertainties, we arrive at our
final result for the spin of A0620--00: $a_* = 0.12 \pm 0.19$~($1
\sigma$)}.  The corresponding radius of the inner disk, which is
uncertain by $12$\%, is just 7\% less than the Schwarzschild value of
$6GM/c^2$.

\section{Discussion}

Any measurement of BH spin is only as good as the theoretical model
behind it.  The continuum-fitting method assumes that the radial
profile of the disk $L(R)$ is given by the analytical form derived by
\citet[][~NT]{nt+1973}.  Because any serious error in the NT model (
e.g., the assumption of vanishing torque at the ISCO) will lead to
large systematic errors in the derived BH spin values, we have mounted
a major effort to scrutinize the NT model.  In \citet{smnt+08}, we
reported a 3D GRMHD simulation of a thin disk ($H/R\sim0.06$) around a
nonspinning BH.  We showed that the angular momentum profile matches
the NT prediction to within $\sim2\%$, indicating that any magnetic
coupling across the ISCO is weak, and we estimated that the additional
disk luminosity due to this weak coupling is only $\sim4\%$.
\citet{nk+09} and \citet{nkh+10} have carried out simulations of
similarly thin disks, but they conclude that deviations from the NT
model are much larger, with the specific angular momentum deviating by
up to 15\%.  In a recent paper~\citep{pmnt+10}, we report simulations
corresponding to a variety of disk thicknesses and BH spins.  For thin
disks with $H/R < 0.07$ we agree with Shafee et al.'s conclusion that
the NT model provides an accurate description of both the angular
momentum profile and the luminosity (see Figures 14 and 15 in Penna et
al.). We suggest that the contrary results obtained by Noble et
al. are because (i) they used for the initial magnetic field in their
simulations a topology with long-range radial coherence, which we
argue is inappropriate for a thin disk, and (ii) they included the
contribution of the corona, even though the corona is generally
believed not to participate in the optically thick thermal emission we
model in the continuum-fitting method.

\begin{figure}[ht]
\plotone{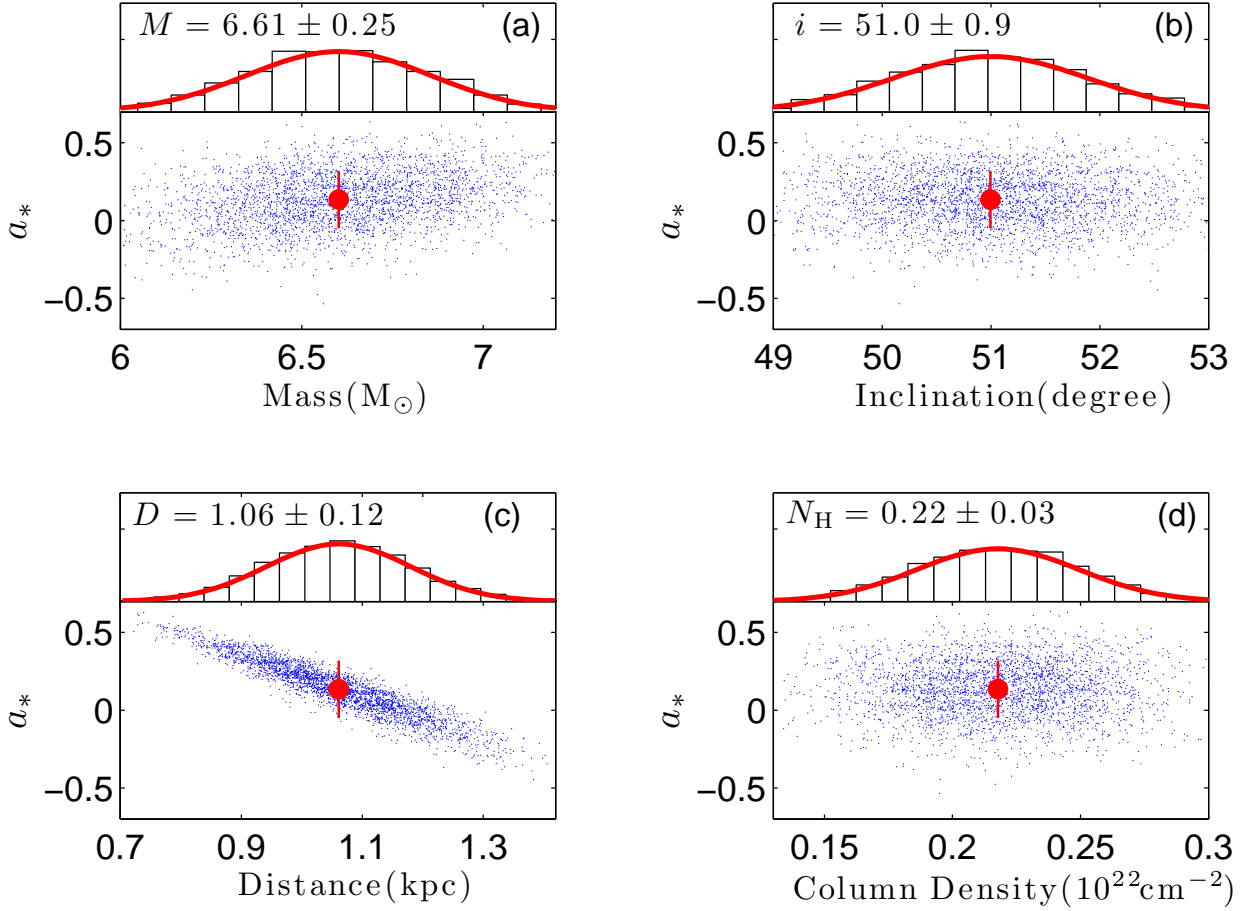}
\caption{Effect on the spin parameter $a_*$ of varying $M$, $i$, $D$ and
  $N_{\rm H}$ for the case $\alpha=0.01$.  (a) The upper panel shows a
  normal distribution for the BH mass $M$ and the lower panel
  shows $a_*$ versus mass $M$ for 3000 sets of parameters drawn at
  random. The central filled circle indicates our estimate of the spin
  $a_{*0}$=$0.135_{-0.179}^{+0.178}$ obtained from these simulations.
  (b--d) Same as panel {\it a} except now for the parameters of
  inclination angle, distance and column density, respectively.}
\label{figuresecond}
\end{figure}

\begin{figure}[ht]
\plotone{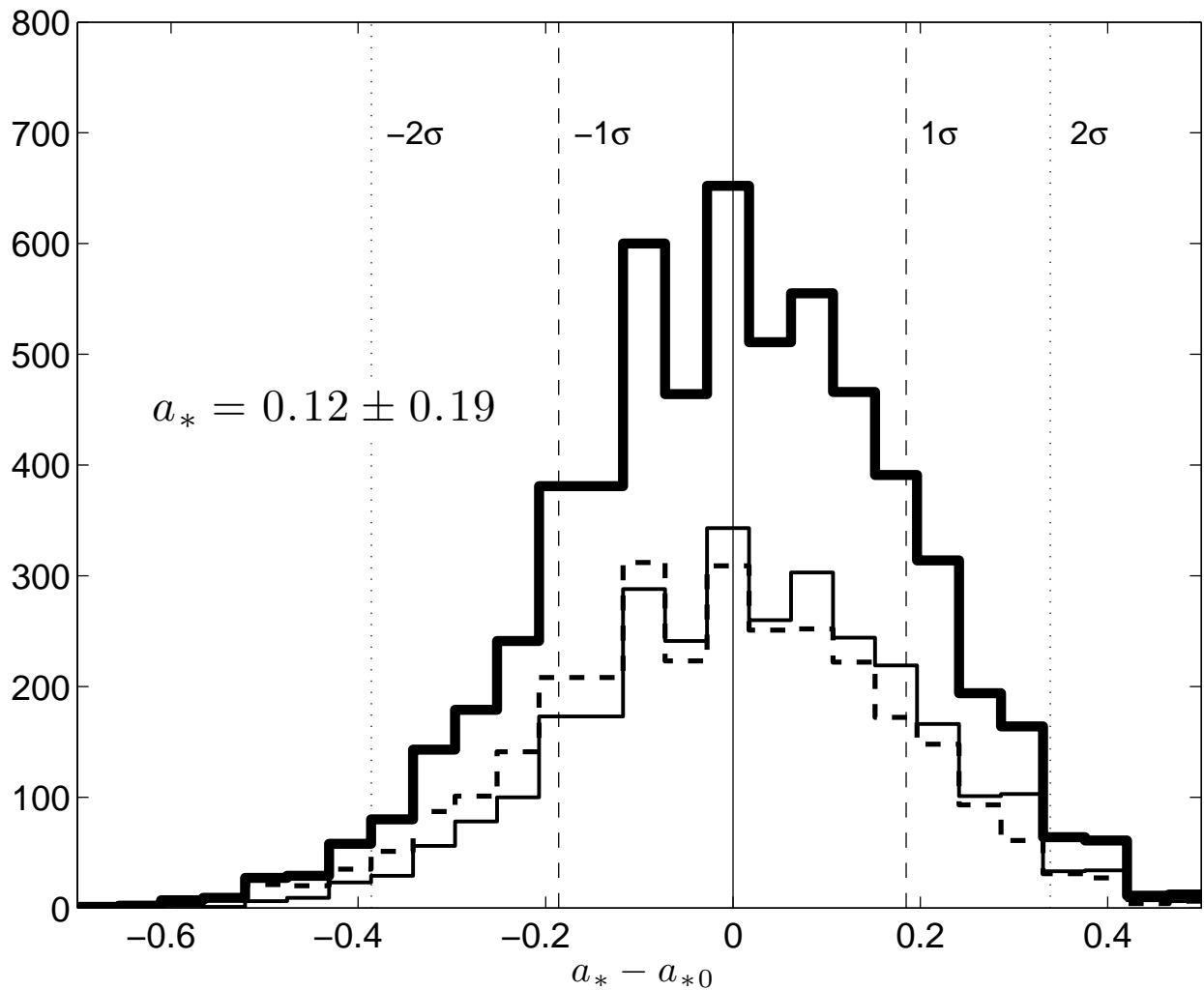}
\caption{Combined error analysis that considers both of our fiducial
  values for the viscosity parameter.  The thin solid line is for
  $\alpha=0.01$ and the dashed line is for $\alpha=0.1$.  The sum of
  these two smaller histograms forms the large histogram.  The
  vertical lines pertain to this latter histogram: The vertical solid
  line indicates the median value of the spin determined by these
  simulations: $a_{*0}=0.12$; the two dashed lines enclose
  68.27$\%$~(1$\sigma$) of the spin values centered on the solid line
  and imply an observational uncertainty of ($-0.19, +0.19$); the two
  dotted lines enclose 95.45$\%$~(2$\sigma$) and imply an uncertainty
  of ($-0.39$, $+0.34$).  The upper and lower limits on the spin at
  the 99.73\% (3$\sigma$) level of confidence are respectively
  $a_*<0.49$ and $a_*$> -0.59.}
\label{figurethird}
\end{figure}

The inner X-ray-emitting portion of a thin accretion disk is
presumably aligned with the spin axis of the BH
\citep[e.g.,][]{lp+07}.  In determining the spin of A0620--00 and most
sources, we assume that the BH's spin is closely aligned with the
orbit vector; a misalignment of more than several degrees would
significantly affect our results (e.g., see
Figure~\ref{figuresecond}$b$).  There is presently no good evidence
for significant misalignments despite two often-cited cases \citep[see
Section 2.2 in][]{nm+2005}.  A recent population-synthesis simulation
study indicates that the majority of BH binaries have relatively small
misalignment angles ($\lesssim10^{\rm \circ}$;
\citeauthor{ftrb+10}~\citeyear{ftrb+10}).  The approved NASA GEMS
polarimetry mission, which is scheduled for a 2014 launch, is expected
to soon provide the capability to determine directly the degree of
alignment to an accuracy of a few degrees \citep{lnm+09,ksjk+09}.

Based on modeling the 1975--76 X-ray and optical light curves,
\citet{sls+08} find a low and consistent value of spin, $a_*\sim0.1$,
for our mass of $M=6.6$~$\msun$ and for a fixed value of $f=1.7$ (see
their Figure\ 7).  However, for this model their fitted value of the
viscosity parameter, $\alpha\approx 0.6$, is significantly higher than
the higher fiducial value of $\alpha=0.1$ that we have adopted
(Section~3; see discussion in Section 5.3 in Gou et al.\ 2009) or the
values of $\alpha\sim0.1-0.4$ based on the ``best observational
evidence'' \citep{kpl+07}.  We are unable to estimate $a_*$ for larger
values of $\alpha$ because the required {\sc bhspec} table models
(Section~3) do not exist.  However, it is clear (e.g., Figure\ 3) that
larger values of $\alpha$ will decrease our estimate of $a_*$.

Recent GRMHD simulations by \citet{frag+09} show that luminous,
geometrically-thick disks with $l\gtrsim 0.6$ (corresponding to
$H/R\gtrsim0.2$; McClintock et al.\ 2006) can falsely appear to host
BHs with low spins if the inner disk is significantly tilted with
respect to the BH spin axis.  It is highly unlikely that this effect
can explain the low spin of A0620--00 because the observed disk luminosity
is low, $l\sim 0.1$ (Table~1), and the disk is correspondingly thin
$H/R\sim0.04$.  In this case, as Fragile notes, one does not expect
the inferred value of spin to be suppressed by the tilted-disk effect
that he describes.

Despite its faint corona and low spin, A0620--00 was a bright
transient radio source in outburst, decaying from $\sim200$~mJy to
$\lesssim10$~mJy in the period from about 12 to 24 days following its
discovery on 3 August 1975.  A reanalysis of these data by
\citet{kfsd+99} indicates multiple jet ejections of initially
optically-thick components.  The authors find that the source was
extended on arcsec scales, and they infer a relativistic expansion
velocity.  Even in quiescence, the radio source was detected at a
level of $\sim0.05$ mJy, indicating the presence of a partially
self-absorbed synchrotron jet \citep{gfmm+06}. If jets are powered by
BH spin, then jet power is likely to increase dramatically with
increasing $a_*$ \citep{mcki+05,tnm+09}. The expected dependence is so
steep that, for $a_*<0.4$, the jet receives more power from the
accretion disk than from the BH \citep{mcki+05}.  Therefore, given the
low spin of A0620--00, it would appear that the jet inferred in this
source was probably driven by the accretion disk, not the BH.  In
closing, we note that a statistical study by \citet{fgr+10}, although
based on data of uneven quality, found no evidence that BH spin powers
jets.

\acknowledgments

JEM acknowledges support from NASA grant NNX08AJ55G and the
Smithsonian Endowment Funds. RN was supported in part by NSF grant
AST-0805832 and NASA grant NNX08AH32G. We thank the following
individuals for help in locating and utilizing the {\it OSO-8} X-ray
spectrum: L.\ Angelini, G.\ Branduardi-Raymont, K.\ Pounds, J.\ Swank
and N.\ White, and we thank J.\ McKinney, S.\ Murray and an anonymous
referee for helpful comments.  This research has made exclusive use of
data obtained from the High Energy Astrophysics Science Archive
Research Center (HEASARC) at NASA/Goddard Space Flight Center.


\begin{deluxetable}{ccccccccc}
\tabletypesize{\normalsize} \tablewidth{0pt}
\tablecaption{\sc kerrbb{\small 2} fit results for A0620--00\tablenotemark{a}}
\tablehead{\colhead{ }&  \colhead{$N_{\rm H}$\tablenotemark{b} } &
\colhead{$\Gamma$} & \colhead{$f_{\rm SC}$} &
\colhead{$a_*$} &
\colhead{$\dot{M}\tablenotemark{c}$}&\colhead{$l\tablenotemark{d}$}&
\colhead{$\chi^2_{\rm \nu}$ }  } \startdata
1 & 2.2&$3.53\pm0.38$&$0.0062\pm0.0030$&$0.135\pm0.029$ &$1.62\pm0.06$&0.109& 0.74
\\ 
2& 1.9\tablenotemark{e}&$3.51\pm0.38$&$0.0060\pm0.0029$&$0.141\pm0.028$&$1.60\pm0.06$&0.109 & 0.74
\\
3& 2.5\tablenotemark{e}&$3.55\pm0.39$&$0.0064\pm0.0032$&$0.126\pm0.029$ &$1.63\pm0.06$&0.109
& 0.74 \\
4 & 2.2&$3.91$\tablenotemark{f}&$0.0099\pm0.0005$&$0.116\pm0.019$ &$1.65\pm0.04$&0.110& 0.75
\\
5 &2.2&$3.15$\tablenotemark{f}&$0.0038\pm0.0002$&$0.155\pm0.025$ &$1.58\pm0.05$&0.108& 0.76
 \\  \enddata
\tablenotetext{a}{$M=6.61~M_{\sun}$, $i=51\fdg0$, $D=1.06$ kpc, and
 $\alpha=0.01$.}
\tablenotetext{b}{Column density is in units of $10^{21}~{\rm
 cm^{-2}}$.}
\tablenotetext{c}{Mass accretion rate is in units of $10^{18}~{\rm
 g~sec^{-1}}$.}
\tablenotetext{d}{$l\equiv L_{\rm bol}(a_*,\dot{M})/L_{\rm
    EDD}$ (see Section 1).}
\tablenotetext{e}{Corresponds to $\rm E(B-V)=0.35\pm0.05$ (see
  Section 3).}
\tablenotetext{f}{Corresponds to $\Gamma=3.53\pm0.38$ from Row 1.}
\label{tablefirst}
\end{deluxetable}



\end{document}